\documentclass[11pt,eqs]{article}
\usepackage{latexsym}

\begin{document}
\title{T-dual-coordinate dependence makes the effective Kalb-Ramond field nontrivial
\thanks{Work supported in part by the Serbian Ministry of Science and
Technological Development, under contract No. 171031.}}
\author{Lj. Davidovi\'c \thanks{e-mail address: ljubica@ipb.ac.rs} and B. Sazdovi\'c
\thanks{e-mail address: sazdovic@ipb.ac.rs}\\
{\it Institute of Physics,}
{\it University of Belgrade,}\\
{\it 11001 Belgrade, P.O.Box 57, Serbia}}
\maketitle

\begin{abstract}
We show  that the fact that the string theory is unoriented
does not necessarily force the
Kalb-Ramond field to vanish. We investigated the
theory of the open string propagating in the weakly curved
background. The effective Kalb-Ramond field
$B^{eff}_{\mu\nu}$, the background field of the effective theory
obtained on the
solution of the boundary conditions, does not depend on the
$\Omega$-even effective coordinate $q$, but on its T-dual
$\tilde{q}$ which is $\Omega$-odd.
This brakes the standard proof that the
term with $B_{\mu\nu}^{eff}$ should vanish.
From the world-sheet equations of motion we
identify $B_{\mu\nu}^{eff}$ with the
torsion potential.
\end{abstract}

It is well known that type I superstring theory
is unoriented (symmetric under world-sheet parity transformation
$\Omega:\sigma\rightarrow -\sigma$).
It can be obtained from type IIB superstring theory as an
$\Omega$-projection.
This procedure eliminates all the states which are odd
under $\Omega$-transformation \cite{GR}-\cite{SN1}.

Generally,
the $\Omega$-odd background fields are
absent in the unoriented theories,
because they come within terms which are integrated out of the
action, as the $\Omega$-odd terms on the
symmetric $\sigma$ interval $[-\pi,\pi]$.
This can be proven using the implicit assumption that
the background fields depend on the
$\Omega$-even coordinates.

We are going to present an exception of this rule,
on the example of the bosonic string.
We will consider the propagation of the open bosonic string,
described by the action  \cite{GR,ZW,BBS,FTFCB}
\begin{equation}\label{eq:action0}
S = \kappa \int_{\Sigma} d^2\xi\Big{[}\frac{1}{2}{\eta}^{\alpha\beta}G_{\mu\nu}(x)
+{\epsilon^{\alpha\beta}}B_{\mu\nu}(x)\Big{]}
\partial_{\alpha}x^{\mu}\partial_{\beta}x^{\nu},
\end{equation}
$(\varepsilon^{01}=-1)$,
where integration goes over two-dimensional world-sheet $\Sigma$
parameterized by the coordinates $\xi^{0}=\tau,\ \xi^{1}=\sigma$ with $\sigma \in [0,\pi]$.
Here $x^{\mu}(\xi),\ \mu=0,1,...,D-1$ are the coordinates of the
D-dimensional space-time, and we use the notation
$\dot{x}=\frac{\partial x}{\partial\tau}$,
$x^\prime=\frac{\partial x}{\partial\sigma}$.

Requirement for the
world-sheet conformal invariance on the quantum level,
produces the restriction on the background fields
which can be represented in a form of
the space-time equations of motion.
We will consider the following particular solution
of these equations,
the {\it weakly curved} background \cite{VS,CSS,DS,DS1}
\begin{eqnarray}\label{eq:gb}
G_{\mu\nu}=const,\quad
B_{\mu\nu}[x]=b_{\mu\nu}+\frac{1}{3}B_{\mu\nu\rho}x^\rho,
\end{eqnarray}
where the parameter $b_{\mu\nu}$ is constant and $B_{\mu\nu\rho}$
is constant and infinitesimally small.

The minimal action principle for the
open string produces
the equation of motion
\begin{equation}\label{eq:motion}
{\ddot{x}}^\mu=x^{\prime\prime\mu}
-2B^\mu_{\ \nu\rho}{\dot{x}}^\nu x^{\prime\rho},
\end{equation}
and the boundary conditions on the string endpoints.
For the Neumann boundary conditions we have
\begin{equation}\label{eq:bonc}
\gamma^{0}_\mu\Big{|}_{\sigma=0,\pi}=0,
\end{equation}
where
\begin{equation}
\gamma^{0}_{\mu}\equiv
\frac{\delta {\cal{L}}}{\delta x^{\prime\mu}}
=G_{\mu\nu}x^{\prime\nu}-2B_{\mu\nu}\dot{x}^\nu.
\end{equation}

We treat the boundary conditions as constraints.
We demand their time consistency and therefrom we obtain the infinite set of constraints
$\gamma^{n}_{\mu}\Big{|}_{\sigma=0,\pi}=0,$
$\gamma^{n+1}_{\mu}\equiv{\dot{\gamma}}^{n}_{\mu},\quad({n\geq 0})$.
Applying the procedure developed in Refs.\cite{DS,DS1},
using the equation of motion (\ref{eq:motion}),
we obtain their explicit form
in the leading order
in $B_{\mu\nu\rho}$.
Instead of working with this infinite set of constraints,
we build a $\sigma$-dependent constraint at
each string end-point. Separating constraint
at $\sigma=0$ in
$\Omega$-symmetric and antisymmetric parts
$\Gamma_{\mu}=\Gamma_{\mu}^{S}+\Gamma_{\mu}^{A}$,
after a long calculation described in \cite{DS,DS1},
its compact form has been obtained
\begin{eqnarray}\label{eq:const}
\Gamma_{\mu}^{S}(\sigma)&=&
G_{\mu\nu}{\bar{q}}^{\prime\nu}
-2b_{\mu\nu}{\dot{q}}^\nu
-\frac{2}{3}B_{\mu\nu\rho}
\Big{[}{\dot{q}}^\nu q^\rho
+\frac{1}{2}{\dot{Q}}^\nu q^{\prime\rho}
\nonumber\\
&+&\frac{3}{2}{\dot{\bar{q}}}^\nu {\bar{q}}^\rho\Big{]}
+2b_{\mu}^{\ \rho}B_{\rho\nu\upsilon}
\Big{[}q^{\prime\nu} {\bar{q}}^\upsilon
+{\dot{Q}}^\nu {\dot{\bar{q}}}^\upsilon\Big{]},
\nonumber\\
\Gamma_{\mu}^{A}(\sigma)&=&
G_{\mu\nu}{\dot{\bar{q}}}^\nu
-2b_{\mu\nu}{q}^{\prime\nu}
-\frac{2}{3}B_{\mu\nu\rho}
\Big{[}q^{\prime\nu}q^\rho
+\frac{1}{2}{\dot{Q}}^\nu{\dot{q}}^\rho
\nonumber\\
&+&\frac{3}{2}{\bar{q}}^{\prime\nu}{\bar{q}}^\rho\Big{]}
+2b_{\mu}^{\ \rho}B_{\rho\nu\upsilon}
\frac{\partial}{\partial\sigma}
\Big{[}{\dot{Q}}^\nu {\bar{q}}^\upsilon\Big{]},
\end{eqnarray}
where we introduced
even and odd coordinate variables
\begin{eqnarray}
q^\mu(\sigma)
\equiv\sum_{n=0}^{\infty}
\frac{\sigma^{2n}}{(2n)!}x^{\mu(2n)}\Big{|}_{\sigma=0},
\nonumber\\
\bar{q}^\mu(\sigma)
\equiv\sum_{n=0}^{\infty}
\frac{\sigma^{2n+1}}{(2n+1)!}x^{\mu(2n+1)}\Big{|}_{\sigma=0},
\end{eqnarray}
and
\begin{equation}\label{eq:qdot}
{\dot{Q}}^\mu(\sigma)=\int_{0}^{\sigma}d\eta {\dot{q}}^\mu(\eta).
\end{equation}

We solve the constraints $\Gamma_{\mu}^{S}(\sigma)=0$,
$\Gamma_{\mu}^{A}(\sigma)=0$
by iteration method.
The solution, in the zero order in the
infinitesimal parameter $B_{\mu\nu\rho}$
is
\begin{eqnarray}\label{eq:solution0}
{\bar{q}}^{\prime\mu}=2b^{\mu}_{\ \nu}{\dot{q}}^\nu,\quad
{\dot{\bar{q}}}^{\mu}=2b^{\mu}_{\ \nu}{q}^{\prime\nu}.
\end{eqnarray}
In the first order in $B_{\mu\nu\rho}$, it becomes
\begin{eqnarray}\label{eq:solution}
{\bar{q}}^{\prime\mu}&=&
2B^{\mu}_{\ \nu}(q){\dot{q}}^\nu
-A^{\mu}_{\ \nu}(\dot{Q}){q}^{\prime\nu},\nonumber\\
{\dot{\bar{q}}}^{\mu}&=&2B^{\mu}_{\ \nu}(q){q}^{\prime\nu}
- A^{\mu}_{\ \nu}(\dot{Q}){\dot{q}}^{\nu},
\end{eqnarray}
where we introduced antisymmetric infinitesimal tensor
\begin{eqnarray}
A_{\mu\nu}(\dot{Q})&=&
h_{\mu\nu}(\dot{Q})
-12b_{\mu}^{\ \rho}h_{\rho\sigma}(\dot{Q})b^{\sigma}_{\ \nu}
\nonumber\\
&-&12h_{\mu\rho}(b\dot{Q})b^{\rho}_{\ \nu}
+12b_{\mu}^{\ \rho}h_{\rho\nu}(b\dot{Q}),
\end{eqnarray}
with $h_{\mu\nu}(x)\equiv \frac{1}{3}B_{\mu\nu\rho}x^\rho$.

If we extend the $\sigma$ domain and demand
$2\pi-$periodicity
of the original variable
$x^\mu(\sigma+2\pi)=x^\mu(\sigma)$,
it can be shown that
the constraints at $\sigma=0$ and $\sigma=\pi$
are equivalent.
Therefore,
the relation (\ref{eq:solution}) solves both constraints at $\sigma=0$ and $\sigma=\pi$.

Substituting
the solution
into the Lagrangian
(\ref{eq:action0})
we obtain the effective one
\begin{eqnarray}\label{eq:efflag}
{\cal{L}}^{eff}&=&
\frac{\kappa}{2}{\dot{q}}^{\mu}G^{eff}_{\mu\nu}(q,\dot{Q}){\dot{q}}^{\nu}
-\frac{\kappa}{2}{q}^{\prime\mu}G^{eff}_{\mu\nu}(q,\dot{Q}){q}^{\prime\nu}
\nonumber\\
&+&2\kappa{q}^{\prime\mu}B^{eff}_{\mu\nu}(q,\dot{Q}){\dot{q}}^{\nu},
\end{eqnarray}
where
\begin{eqnarray}
G^{eff}_{\mu\nu}(q,\dot{Q})&=&
G^{E}_{\mu\nu}(q+2b\dot{Q})
+4[b^{2}A(\dot{Q})-A(\dot{Q})b^{2}]_{\mu\nu},
\nonumber\\
B^{eff}_{\mu\nu}(q,\dot{Q})&=&
[h(2b\dot{Q})+4bh(2b\dot{Q})b]_{\mu\nu}
-B_{\mu}^{\ \rho}(q)G_{\rho\nu}^{E}(q),
\nonumber\\
\end{eqnarray}
and
\begin{equation}
G^{E}_{\mu\nu}(x)\equiv G_{\mu\nu}-4B_{\mu\rho}(x)(G^{-1})^{\rho\sigma}B_{\sigma\nu}(x)
\end{equation}
is the open string metric.

Because our basic variable $q^\mu (\sigma)$ contains only even
powers of $\sigma$,
it is convenient to regard it as the even function
$q^\mu (-\sigma)=q^\mu (\sigma)$ on the interval $\sigma \in [-\pi,\pi]$. 
Therefore, we will consider the action
$S^{eff} = \int d\tau \int_{-\pi}^\pi d\sigma {\cal{L}}^{eff}$,
which makes our effective theory
unoriented closed string theory.
Consequently, the terms of the effective metric which depend on
$\dot{Q}$ and the terms of effective Kalb-Ramond field which depend
on $q$ will disappear and we obtain
\begin{equation}\label{eq:aceff}
S^{eff} = \kappa \int_{\Sigma_1} d^2\xi   \Big[\frac{1}{2}\eta^{\alpha\beta}G^{eff}_{\mu\nu}(q)
+{\epsilon^{\alpha\beta}}B^{eff}_{\mu\nu}(2 b \dot{Q})\Big]
\partial_{\alpha}q^{\mu}\partial_{\beta}q^{\nu},
\end{equation}
where $\Sigma_1$ marks the changed sigma domain $\sigma \in [-\pi, \pi]$.
The effective background fields are equal to
\begin{eqnarray}
G^{eff}_{\mu\nu}(q)= G^{E}_{\mu\nu}(q) \, ,  \quad
B^{eff}_{\mu\nu}(2b \dot{Q})= -\frac{\kappa}{2} [g \Delta\theta (2b \dot{Q})g]_{\mu\nu},
\nonumber\\
\end{eqnarray}
where $\Delta\theta$ is the infinitesimal part of
the so called non-commutativity parameter
\begin{equation}
\theta^{\mu\nu}=-\frac{2}{\kappa}\Big{[}G^{-1}_{E}BG^{-1}\Big{]}^{\mu\nu},
\end{equation}
and $g_{\mu\nu}=G^{E}_{\mu\nu}(0)$ is the constant part of the
effective metric.

There are two unexpected results here. The first one is the appearance
of the non-trivial Kalb-Ramond field $B^{eff}_{\mu\nu}$ and the
second one is the fact that it does not depend on the coordinate
$q^\mu$ but on
${\dot{Q}}^\mu(\sigma)=\int_{0}^{\sigma}d\eta{\dot{q}}^\mu(\eta)$.
We are going to offer an explanation and the interpretation of these
results.

It is well known that the theory of the unoriented closed
string (which is just our effective theory)
should not contain the
Kalb-Ramond field.
Let us first present the standard reasons,
for this statement
\cite{ZW,SN}.
The effective Kalb-Ramond field appears in the 
effective action within the term 
$B^{eff}_{\mu\nu}{\dot{q}}^\mu q^{\prime\nu}$.
So, as ${\dot{q}}^\mu$ is $\Omega$-even and $q^{\prime\mu}$ $\Omega$-odd variable,
if the Kalb-Ramond field depends on the $\Omega$-even variable,
this term does not contribute,
because it disappears after integration over the symmetric interval
$[-\pi,\pi]$.

What is different in our case?
It is the fact that the effective Kalb-Ramond field does not depend
on the effective coordinate $q^\mu$ ($\Omega$-even)
but on the integral of the
$\tau$-derivative of the effective coordinate
${\dot{Q}}_\mu(\sigma)=\int_{0}^{\sigma}d\eta {\dot{q}}_\mu(\eta)$
($\Omega$-odd).
Since, $B_{\mu\nu}^{eff}(2b\dot{Q})$ is linear in $\dot{Q}^\mu$
it means that the effective Kalb-Ramond field
is odd under $\sigma$-parity transformation
$\Omega B_{\mu\nu}^{eff}[2b\dot{Q}(\sigma)]=-B_{\mu\nu}^{eff}[2b\dot{Q}(\sigma)]$.
This minus sign changes the situation, because the term in the action with the
effective Kalb-Ramond field becomes $\Omega$-even,
and this fact allows its survival.

What can the interpretation of ${\dot{Q}}^\mu$ be?
First, note that ${\dot{Q}}^\mu$
appears as an argument of $B^{eff}_{\mu\nu}$ only, which is the
infinitesimal of the first order. So, it is enough to consider
${\dot{Q}}^\mu$ up to the zero order. The zero order equation of
motion for $q^\mu$ is just $\partial_{+}\partial_{-}q^\mu=0$, and
consequently the solution has the form
$q^\mu(\sigma)=f^\mu(\sigma^{+})+g^\mu(\sigma^{-})$,
with $\sigma^{\pm}=\tau\pm\sigma$.
The $\Omega$-evenness of the variable
$q^\mu$,
$q^\mu(-\sigma)=q^\mu(\sigma)$,
implies $f(\sigma)=g(\sigma)$
and we obtain
\begin{equation}
q^\mu(\sigma)=f^\mu(\sigma^{+})+f^\mu(\sigma^{-}).
\end{equation}
From the properties $\partial_{\pm}f^{\mu}(\sigma^{\mp})=0$, we have
${\dot{f}}^{\mu}(\sigma^{\pm})=\pm f^{\prime\mu}(\sigma^{\pm})$.
Therefore, $ {\dot{q}}^\mu(\sigma)=f^{\prime\mu}(\sigma^{+})
-f^{\prime\mu}(\sigma^{-}), $ and consequently
\begin{equation}
{\dot{Q}}^\mu(\tau,\sigma)= f^{\mu}(\sigma^{+})
-f^{\mu}(\sigma^{-}) \equiv {\tilde{q}}^\mu(\tau,\sigma),
\end{equation}
where ${\tilde{q}}^\mu(\tau,\sigma)$ is T-dual mapping of the
effective coordinate $q^\mu(\tau,\sigma)$ (see for example
(17.76) of Ref. \cite{ZW}
or
eq. (6.17) of Ref.\cite{BBS}).
Note that T-dual coordinate $\tilde{q}$ is
$\Omega$-odd variable, because
$\tilde{q}(-\sigma)=-\tilde{q}(\sigma)$.
So, in the effective theory with
dynamical variable
$q^\mu$, the effective metric depends on the coordinate
$q^\mu$ and the effective Kalb-Ramond field on its T-dual
 $\tilde{q}^\mu$.

The question arises, why do the background fields depend on the different arguments?
To understand this,
it is enough to consider the zero order solution
of the boundary conditions,
because the
arguments of the background fields appear only in the infinitesimally
small terms. Let us rewrite the solution
of the constraints (\ref{eq:solution0}),
in the form
\begin{equation}\label{eq:xqq}
x^\mu(\sigma)=q^\mu(\sigma)+2(G^{-1}b)^{\mu}_{\ \nu} {\tilde{q}}^\nu.
\end{equation}
The effective metric depends
on the first ($\Omega$-even) part of $x^\mu$
and the effective Kalb-Ramond field
on the second ($\Omega$-odd) part of $x^\mu$.
So, we can formally rewrite the effective action
(\ref{eq:aceff}) in the form
\begin{eqnarray}\label{eq:leff}
S^{eff} = \kappa \int_{\Sigma_1} d^2\xi \Big{[}
\frac{1}{2}\eta^{\alpha\beta} G^{eff}_{\mu\nu}(x)
+ \epsilon^{\alpha\beta} B^{eff}_{\mu\nu}(x)
\Big{]}
\partial_\alpha q^\mu \partial_\beta q^\nu,
\nonumber\\
\end{eqnarray}
because
$\eta^{\alpha\beta} [G^{eff}_{\mu\nu}(2b{\tilde{q}})
-G^{eff}_{\mu\nu}(0)]
\partial_\alpha q^\mu \partial_\beta q^\nu$
and
$\epsilon^{\alpha\beta} B_{\mu\nu}^{eff}(q)
\partial_\alpha q^\mu \partial_\beta q^\nu$
do not contribute being the $\Omega$-odd terms.
Therefore, formally, both background fields
depend on the same argument,
the solution for the
initial coordinate $x^\mu$ expressed as in (\ref{eq:xqq}), in terms of the effective one $q^\mu$
and its T-dual ${\tilde{q}}^\mu$.

The appearance  of the T-dual effective coordinate
${\tilde{q}}^\mu = \dot{Q}^\mu$ in the Lagrangian (\ref{eq:leff})
suggests that we have obtained
the nonlocal theory. It is interesting to find the equation of motion
of the theory of this type,
when $G^{eff}_{\mu\nu}$ is arbitrary function of $q^\mu$ and
$B^{eff}_{\mu\nu}$ consists of the odd powers in $\Omega$-odd
variable $\bar{q}=2b{\tilde{q}}$.

Variation with respect to $q^\mu$
produces world-sheet equation of motion
\begin{eqnarray}
\eta^{\alpha\beta}\ {}^{0}D_{\beta}\partial_\alpha q^\mu
= 2\kappa \theta^{\mu\nu}_{0}
\epsilon^{\alpha\beta}\int_{0}^{\sigma}d\eta
\frac{\partial}{\partial\tau}
\Big{[}
\frac{\partial B^{eff}_{\rho\sigma}(\bar{q})}{\partial\bar{q}^{\nu}}
\partial_{\alpha}q^\rho\partial_{\beta}q^\sigma
\Big{]},
\nonumber\\
\end{eqnarray}
where
${}^{0}D_{\alpha}V^\mu=\partial_\alpha q^\nu {}^{0}D_{\nu}V^\mu$
is the covariant derivative in the world-sheet direction.
${ }^{0}D_\mu$ is generalized covariant derivative
$^{0}D_\mu V^\nu=\partial_\mu V^\nu+
{ }^{0}\Gamma^{\nu}_{\rho\mu}V^\rho$,
with the generalized connection
\begin{equation}
{ }^{0}\Gamma^{\mu}_{\rho\sigma}=(\Gamma_{eff})^{\mu}_{\rho\sigma}
+\frac{1}{2}{ }^{0}K^\mu_{\ \rho\sigma}.
\end{equation}
It consists of the Christoffel connection $(\Gamma_{eff})^{\mu}_{\rho\sigma}$
and contorsion ${ }^{0}K_{\mu\rho\sigma}=\frac{1}{2}{ }^{0}T_{\{\sigma\mu\rho\}}$
in terms of torsion
\begin{equation}\label{eq:qstr}
{ }^{0}T_{\mu\rho\sigma}=
4b_{\rho}^{\ \nu}
\frac{\partial}{\partial {\bar{q}}^\nu}
 B^{eff}_{\mu\sigma}.
\end{equation}
Here $\{\mu\nu\rho\}=\nu\rho\mu+\rho\mu\nu-\mu\nu\rho$
is Schouten bracket.

Therefore, the Kalb-Ramond field is related to the
torsion potential.
This
is in accordance
with the
usual interpretation
of the Kalb-Ramond field $B_{\mu\nu}$
in the low energy string theory
as the non-Rimannian theory \cite{SP}.
In our case torsion is
infinitesimally small constant.

Let us summarize.
In our approach the initial theory represents an oriented open string theory.
The theory obtained on the solution of the boundary conditions,
to which we refer as an effective theory, is an unoriented
closed string theory, because it is symmetric under $\sigma$-parity
transformation, $\Omega :\sigma\rightarrow -\sigma$ and satisfies the boundary condition
$q^\mu(\sigma=-\pi)=q^\mu(\sigma=\pi)$.

The complete transition from the original theory
(\ref{eq:action0}) to the effective theory (\ref{eq:aceff})
consists of
\begin{itemize}
\item[1.]{ {\it the dynamical variable} transition
\begin{equation}\label{eq:vt}
x^\mu\rightarrow q^\mu,
\end{equation}
} \item[2.]{ {\it the background field} transition
\begin{eqnarray}\label{eq:ggbb}
G_{\mu\nu}\rightarrow G_{\mu\nu}^{eff}(q), \qquad
B_{\mu\nu}(x)\rightarrow B^{eff}_{\mu\nu}(2b{\tilde{q}}) \, .
\end{eqnarray}
}
\end{itemize}
The effective action (\ref{eq:aceff}) describes the propagation
of the effective unoriented closed string
in the effective background (\ref{eq:ggbb}).

When both background fields of the initial theory, $G_{\mu\nu}$
and $B_{\mu\nu}$, are constant the effective background has
constant effective metric $g_{\mu\nu}=G^{E}_{\mu\nu}(0)$ and
zero Kalb-Ramond field. Infinitesimal correction of the initial
antisymmetric field $B_{\mu\nu}$, linear in coordinate $x^\mu$
produces infinitesimal correction of the effective metric linear
in effective coordinates $q^\mu$ and turns the infinitesimal part of the
Kalb-Ramond field into the
T-dual effective coordinate ${\tilde{q}}^\mu$ dependent.
This fact makes the term with $B^{eff}_{\mu\nu}$, $\Omega$-even 
and allows its existence in the unoriented string theory.

The non-commutativity properties of the open string
in the weakly curved background
are
considered in Refs. \cite{DS,DS1}.


\begin{thebibliography}{99}
\bibitem{GR} M. B. Green {\sl et al.} {\it Superstring
Theory}, Cambridge University Press, 1987;
J. Polchinski, {\it String theory}, Cambridge University Press, 1998.

\bibitem{ZW} Zwiebach, {\it A First Course in String Theory}, Cambridge University Press, 2004.

\bibitem{BBS} K. Backer {\sl et al.} {\it String Theory and M-theory}, Cambridge University Press, 2007.

\bibitem{SN1} B. Nikoli\'c {\sl et al.} Phys. Lett.
{\bf B 666} (2008) 400; B. ~Nikoli\'c {\sl et al.}
Nucl. Phys. {\bf B 836} (2010) 100.
B. ~Nikoli\'c {\sl et al.} JHEP, {\bf 08} (2010) 037.

\bibitem{FTFCB} E. S. Fradkin {\sl et al.} Phys. Lett. {\bf B 158} (1985) 316;
{\it Nucl. Phys.} {\bf B 261} (1985) 1.
C. G. Callan {\sl et al.} Nucl. Phys.
{\bf B 262} (1985) 593;
T. Banks {\sl et al.} Nucl. Phys. {\bf B 277} (1986) 67;
N. Seiberg {\sl et al.} JHEP {\bf 09} (1999) 032.

\bibitem{VS} V.Schomerus, {\sl Lectures of Branes in Curved Backgrounds},
Class. Quant. Grav. {\bf 19} (2002) 5781.

\bibitem{CSS} L.Cornalba {\sl et al.} Commun.Math.Phys. {\bf 225} (2002) 33;
M. Herbst {\sl et al.} JHEP {\bf 09} (2001) 014;
A.Yu. Alekseev {\sl et al.} JHEP {\bf 09} (1999) 023.

\bibitem{DS} Lj. Davidovi\'c {\sl et al.} {\it Phys. Rev.} {\bf D 83} (2011) 066014.

\bibitem{DS1} Lj. Davidovi\'c {\sl et al.},
{\it Momenta dependent non-commutativity parameters and background fields},
 in preparation.

\bibitem{SN} B. ~Sazdovi\'c, {\it Eur. Phys. J} {\bf C44} (2005) 599; B. ~Nikoli\'c {\sl et al.}, Phys. Rev.
{\bf D74} (2006) 045024; B. ~Nikoli\'c {\sl et al.} {\it Phys. Rev.} {\bf D75} (2007) 085011;
B. ~Nikoli\'c {\sl et al.} Adv. Theor. Math. Phys. 14 (2010) 1.

\bibitem{SP} T.L.Curtright {\sl et al.} {\it Phys. Rev. Lett.} {\bf 53} (1984) 1799;
{\it Phys. Rev.} {\bf B 162}, (1985) 345;
D.S.Popovi\'c {\sl et al.} {\it Eur. Phys. J} {\bf C 50} (2007) 683.

\end{thebibliography}
\end{document}